\documentstyle[aps,prb,preprint]{revtex}
\begin{document}
\draft
\title{Comment on  ``Critical current density from magnetization hysteresis data using the critical-state model'' }
\author{P. Chaddah}
\address{Low Temperature Physics Laboratory,\\
Centre for Advanced Technology,\\ Indore 452013, India}
\date{\today }
\maketitle
\begin{abstract}
A recent paper [Physical Review {\bf B 64} 014508 (2001)] claims to present an exact method to extract the critical current density J$_C$ from the M-H hysteresis curve. We show that this claim appears unjustified.

\end{abstract}
The hysteresis in the isothermal M-H curves of superconductors was first related to the critical current density J$_C$ by Bean's critical state model \cite{1}. Fietz and Webb \cite{2} showed that the errors involved 
in extracting the field-dependent J$_C$(B) from these M-H curves are of order (d$^2$J$_C$/dB$^2$). The applicability of this method of extracting J$_C$(B) has been studied in great detail \cite{3,4} after the discovery of high-T$_C$ superconductors. The recent claim \cite{5} of an `exact method' of obtaining the same information from the same data, consequently, needs to be examined.

The `exact method' of reference[5] relates, in its equations (7) and (8), the derivatives of the forward and reverse envelope M-H curves to J$_C$(B$_a$), and to the difference between the applied field B$_a$ and the field B$_{C+}$ (or B$_{C-}$ ) at the centre of the sample during the increasing (or decreasing) field cycle. It then presents a method (see figures 2 and 3 of reference[5]) to relate the central fields B$_{C+}$ (or B$_{C-}$) to B$_a$ via the measured M-H curves. All these results are obtained for a sample in the shape of an infinite slab parallel to the applied field.

Analytical solutions of Bean's critical state model, for various J$_C$(B), are known for infinite cylinders (of various cross-sections) in parallel field geometry, when the sample demagnetization factor N=0 .\cite{3,6,7} It is known that while field profiles B(x) depend on J$_C$(B) and not on the shape of the cylinder's cross-section, the value of the magnetization M is dictated by the cylinder's cross-section .\cite{6,7} While equations (5) and (6) of reference[5] are valid for all N=0 cylinders, the derivation leading to equations (7) and (8) breaks down. This is simply because the volume element at x depends on the shape of the cylinder's cross-section, resulting in different x-dependent weight-functions within the integral sign. Equations(11) of reference[3] give explicit analytic expressions for the M-H curves of circular cylinders when J$_C$(B) follows the exponential model J$_C$(B) = J$_C$(0)exp(-B/H$_0$). It is also known \cite{6} for the exponential model that B$_{C+}$ = H$_0$ log[exp(B$_a$/H$_0$) - (H$^*$/H$_0$)] and that B$_{C-}$ = H$_0$ log[exp(B$_a$/H$_0$) + (H$^*$/H$_0$)]. It is straightforward to confirm that equations (7) and (8) of reference[5] are not satisfied.

The argument leading to figure 3(b) of reference[5] is based on the relation (depicted in figure 2 thereof) that the total flux contained in the sample is the same in the field-increasing case when the applied field is B$_a$, as in the field-decreasing case when the applied field is B$_{C+}$ (corresponding to that B$_a$). It follows from the results in reference[3] that this is not true for circular cylinders for J$_C$(B) following the exponential or the Kim-Anderson models. It follows from reference[1] that this is not true for a circular cylinder even when J$_C$(B) is independent of B. It is straightforward to see that this result also breaks down for cylinders with other cross-sections \cite{6}.

We assert that the method of reference[5] is applicable only for an infinite slab in parallel field, and is not valid for any other sample shape. The conventional method \cite{1,2,3,4} for extracting J$_C$(B) from isothermal M-H hysteresis curves is valid for various shapes, and involves errors whose upper bounds are known.

Finally, reference[5] claims to show that the shape of the hysteresis bubble identifying the peak-effect in J$_C$(B) is asymmetric about the peak field B$_P$. This is shown using the input that J$_C$(B) decays as the magnitude of (B-B$_P$) rises. While this result has been shown earlier explicitly with reference to the peak effect (see figure 2b of reference[7]), the calculation is identical to the special case of B$_P$=0. This asymmetry in the `central peak position' has been studied extensively in recent years \cite{8}.

\end{document}